%% file: main.tex
\documentclass[reprint,aps,prl,doublecolumn,amsmath,amssymb]{revtex4-2}

\usepackage{amsmath}
\usepackage{amsfonts}
\usepackage{amsthm}
\usepackage{graphics}
\usepackage{graphicx}
\usepackage{amssymb}
\usepackage {mathrsfs}
\usepackage{braket}
\usepackage{color}
\usepackage{float}
\usepackage[abs,percent]{overpic}
\usepackage[usenames,dvipsnames]{xcolor}
\usepackage{dcolumn}
\usepackage{bm}
\usepackage{xcolor}
\usepackage[hidelinks]{hyperref}
\hypersetup{colorlinks=true, linkcolor=black, citecolor=black, urlcolor=blue}
\usepackage{adjustbox}
\usepackage{url}
\usepackage{ulem}

\usepackage[flushleft]{threeparttable}

\begin{document}
\author{Mohamed Hatifi}
\email[Corresponding author: ]{mohamed.hatifi@oist.jp}
\author{Anshuman Nayak}
\email[Corresponding author: ]{anshuman.nayak@oist.jp}
\author{Jason Twamley}
\affiliation{Quantum Machines Unit, Okinawa Institute of Science and Technology Graduate University, \\Onna, Okinawa 904-0495, Japan} 

\title{Spin-mechanical thermal machines} 


\begin{abstract}

We introduce a method to construct a quantum battery and a quantum Otto heat engine using a Nitrogen-Vacancy (NV) center spin coupled to a mechanical oscillator in a highly detuned regime. By precisely controlling the NV spin, we enable efficient energy exchange despite significant detuning, challenging conventional assumptions. This leads to a robust mechanical quantum battery and a cooling scheme driving the oscillator toward its ground state. Leveraging this, we implement a quantum Otto engine that produces mechanical work at maximum efficiency without decoupling from the hot bath, paving the way for practical quantum thermal machines.
\end{abstract}
\date{\today}
\maketitle
\paragraph{Introduction:-}
The intersection of thermodynamics and quantum mechanics has opened new fields of research for understanding energy transfer at the quantum level \cite{schulman1999,lostaglio2015,uzdin2015, strasberg2017,josefsson2018,chen2019,bera2021,Dynamical2022,Enhancement2023,feyisa2024}. In classical thermodynamics \cite{callen1991}, heat refers to a form of energy that, when transferred, produces entropy. This energy transfer generally occurs through thermal conduction, convection, or radiation and is fundamental for maintaining thermal equilibrium in a system. On the other hand, work is defined as the transfer of energy that does not increase entropy, usually when the system moves or changes shape. Sadi Carnot demonstrated two centuries ago \cite{carnot1824}, in 1824,  that work can be extracted via reversible cycles where a working fluid alternates contact between hot and cold reservoirs. 
Miniaturizing thermodynamic machines towards the quantum regime may have 
practical applications 
in energy extraction and storage such as quantum batteries 
\cite{downing2023,qu2023,zhang2024,yang2024,gangwar2024} and quantum heat engines (QHEs) \cite{josefsson2018,klatzow2019,beyer2019,ono2020,kamimura2022,bera2022,ji2022}. Extracting work from quantum thermal engines has been studied 
in various quantum systems theoretically and experimentally \cite{josefsson2018,klatzow2019,beyer2019,ono2020,bera2021,fujiwara2021,kamimura2022,bera2022,ji2022,boubakour2023,laskar2024,kodama2024}. These include single qubits coupled to thermal baths \cite{yang2014,josefsson2018,barontini2019,barontini2019b,razavian2019,ono2020,ma2024}, quantum resonators interacting with reservoirs \cite{bennett2020,boubakour2023,rodin2024,rodin2024b}, and systems where the statistics of identical bosons or fermions are manipulated \cite{li2011,chen2018,myers2020,koch2023}. 
Despite these advancements, achieving total control over the interaction between quantum systems and their environments remains a significant challenge.  Controlling energy flow at the quantum scale promises significant advancements in nanotechnology \cite{bera2021,fujiwara2021,laskar2024,kodama2024} and materials science \cite{azimi2014,tulzer2020}, particularly in quantum thermometry \cite{levy2020,fujiwara2021,aybar2022,zhao2023,abrahams2023}. Significant progress has been made in manipulating spin-mechanical systems for applications in quantum sensing \cite{riviere2022,pellet2023}, simulations \cite{monroe2021,dobrzyniecki2023}, and energy transduction \cite{bejarano2024}. However, a large disparity typically exists between the bare motional and spin frequencies, which are trying to be addressed by dressing the spin transitions via driving fields \cite{Rabl2010Cooling,Cirio2012}.
\begin{figure}[h!]
    \centering
    \includegraphics[width=1\linewidth]{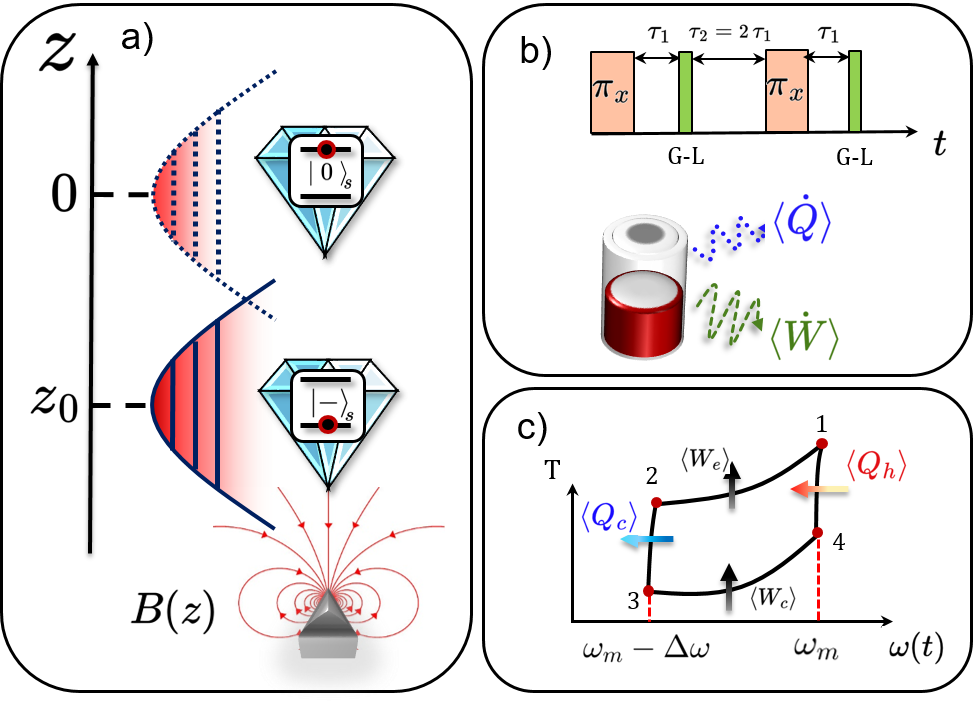}
    \caption{{\bf (a)} {\it Experimental Setup:} A harmonically trapped nanodiamond containing an NV center spin is subjected to a gradient magnetic field \( B(z) = B_0 - G_z z \). In the high magnetic-field limit, the NV center approximates a spin-1/2 system with states \( \ket{0}_s \) and \( \ket{-1}_s \). The spin-mechanical coupling \( \sigma_z (a + a^\dagger) \) induces a spin-dependent shift in the trap position, enabling the creation of two thermal machines illustrated in (b) and (c).
    {\bf (b)} {\it Quantum Battery:} Energy is stored in the high-quality mechanical resonator due to the spin-dependent coupling and the long mechanical coherence time (\( \gamma_m \ll \Gamma_1, \Gamma_2 \)). The protocol alternates between transverse spin pulses, free evolution, and spin resetting via optical means, effectively transferring and storing energy in the mechanical mode.
    {\bf(c)} {\it Quantum Heat Engine:} By adiabatically modulating the mechanical frequency between \( \omega_m - \Delta\omega \) and \( \omega_m \) and precisely tuning the spin pulses, an effective cold reservoir is engineered. This allows energy extraction from the mechanical resonator using a single thermal bath at ambient temperature.}\label{fig:Qbat}
\end{figure}
\\
In this work, we propose a nanoscopic quantum engine using a high-Q nanodiamond trapped in high vacuum \cite{Delord2016, OBrien2019, Tian2024FeedbackPlate}. The center-of-mass (CoM) motion of the particle, with frequencies around $10$--$100$ Hz, serves as the working fluid. Embedded within the nanodiamond is a spin system operating at GHz frequencies. We engineer spin-mechanical coupling via magnetic field gradients to achieve coherent exchange between the spin and the particle's motion. Crucially, the spin can be {\it reset} through optical pulses — a spin polarization process that irreversibly reduces the spin's entropy.
By coupling the spin reset with spin-motional interactions, we effectively create an {\it internally generated} thermal bath for the working fluid, with an adjustable effective temperature that can be lower or higher than the single bath temperature of the mechanical oscillator. Using this controllable spin-mechanical bath, we demonstrate how energy can be pumped into and extracted from the long-lived mechanical motion of the trapped particle, effectively functioning as a quantum battery. Furthermore, by periodically modulating the trapping strength, we construct a quantum engine where the COM motion is the working fluid, the hot bath is the ambient environment, and the cold bath is engineered via sequences of spin polarization and spin-mechanical coupling. Remarkably, this engine can achieve efficiencies surpassing the Curzon-Ahlborn limit and reach Carnot efficiency without the need for quasi-static operation.
\paragraph{The model:-}
We consider a nanodiamond containing an NV center whose position is harmonically trapped along $\Vec{e}_z$ at a low frequency $\omega_m$. The center of mass motion of the NV spin is quantized and can be described  by the Hamiltonian
\begin{equation}\label{Hmech0}
    \mathcal{H}_{\text{m}}/\hbar= \omega_m a^{\dagger}a\;\;.
\end{equation}
The NV spin experiences a static external magnetic field of the form $\Vec{B}=(B_0-G_z z)\Vec{e}_z$, (generated by a nearby sharp permenant magnetic tip), where $B_0$ is a constant magnetic field and $G_zz$ is the field due to the magnetic gradient $G_z$ along $\Vec{e}_z$ at the spatial position $z$ relative to the magnet tip.
We further assume the NV axis to coincide with the direction of the magnetic field, i.e., along $\Vec{e}_z$. The spin Hamiltonian can be described as follows:
\begin{equation}
    \mathcal{H}_{\text{spin}}/\hbar= D~S_z^2+\gamma \Vec{B}\cdot\Vec{S}\;\;,
\end{equation}
with $D$, the zero-field splitting of the electronic energy levels of the NV center, $\vec{S}$ the spin-1 vector operator with components $(S_x, S_y, S_z)$, and $\gamma$ the gyromagnetic ratio of the electron. This explicitly takes the form
\begin{equation}\label{Hspin}
    \mathcal{H}_{\text{spin}}/\hbar= D~S_z^2+\gamma B_0S_z-\gamma G_zzS_z\;\;.
\end{equation}
The center of mass position $z$ can be further expressed in terms of ladder operators as:
    $z=\left(a+a^{\dagger}\right)\times z_{zpf}/\sqrt{2}$
where $z_{zpf}=\sqrt{\hbar/(m\omega_m)}$ is the zero point fluctuation and $m$ is the mass of the nanodiamond. The last term of \eqref{Hspin} couples the spin and the mechanics through the term $z\,S_z$. 
In the case where $B_0>1024\,{\rm G}$, (the magnetic field where the $m_S=0$ and $m_S=-1$ levels cross), if we consider $\Omega_{L}\ll2D-\delta$, where $\delta=D-\gamma B_0-\omega_L$, the state $m_s=+1$ can be ignored, and the Hamiltonian in the rotating frame is expressed in terms of a two-level system $\ket{S}\equiv \ket{m_S},\;\;S=\{-1,0\}$, coupled to the mechanical mode through
\begin{equation}\label{Hspin2}
    \mathcal{H}/\hbar=\omega_m a^{\dagger}a +\frac{\Delta}{2}\left(1-\sigma_z\right)-\frac{g}{2}\,\left(a+a^{\dagger}\right)\,\left(1-\sigma_z\right)+\Omega\sigma_x
\end{equation}
where $\Delta=\omega_L-\omega_0$, $g=\gamma G_z\,z_{\text{zpf}}$, and $\omega_0=\gamma B_0-D$.

When the spin is in the state $\ket{0}$, the system feels no spin-mechanical force and relaxes slowly to its associated equilibrium position, which we assume to have $\langle z\rangle=0$. In contrast, in the state $\ket{-1}$, the system's equilibrium position moves to $z_0=\gamma G_z/m \omega_m^2$, as depicted in Fig \ref{fig:Qbat}(a). 

 In addition, we consider the spin to be driven using an amplitude modulated microwave field, which can be represented by
\begin{equation}
    \mathcal{H}_D/\hbar=\Omega(t)\left(\sigma_+\,e^{-i\omega_L t}+\sigma_-\,e^{i\omega_L t}\right)
\end{equation}
where $\Omega=\Omega_{L}f(t)$. $\Omega_{L}$ represents the maximal amplitude of the drive and we choose a binary dependence for $f(t)=\{0,1\}$, and thus arbitrary pulse trains can be fashioned by a suitable choice of $f(t)$.

We describe the evolution of the full spin-mechanical system with a density matrix $\rho$, which evolves through the master equation defined by
\begin{equation}\label{be}
    \frac{d}{dt}\rho=\frac{1}{i\hbar}\,\left[\mathcal{H},\rho \right]+\sum_k{\cal L}_{X_k}[x_k]\rho+\sum_k{\cal L}_{Y_k}[y_k]\rho\;\;,
\end{equation}
\begin{align}
    {\cal L}_{X_k}[x_k]\rho&=\frac{\gamma_{X_k}}{2}\left[(\bar{n}_{X_k}+1){\cal L}[x_k]\rho+\bar{n}_{X_k}{\cal L}[x_k^{\dagger}]\rho\right] \;\;,\\
    {\cal L}_{Y_k}[y_k]\rho&=\frac{\gamma_{Y_k}}{2}{\cal L}[y_k]\rho\;\;,
\end{align}
where $x_k=\{ a, \sigma_- \}$, with rates $\gamma_{X_k}=\{ \gamma_m, \gamma_1 \}$, describes the coupling of the motion and spin to their thermal baths, and $y_k=\{ \sigma_z, \sigma_- \}$, with rates $\gamma_{Y_k}=\{ \gamma_2, \Gamma_{GL} \}$, describe spin dephasing and the process of spin polarization by the green light illumination. Here $\gamma_m$ is the mechanical damping rate, and $\bar{n}_{X_k}=(\exp[\hbar\omega_k/k_BT_k]-1)^{-1}$, is the mean thermal occupation of the mechanical(spin) mode $(\omega_m(\omega_0)$ at temperature $T_k$ of the thermal environments and $k_B$ is the Boltzmann constant.
The rate $\gamma_1$ represents the spin amplitude damping, $\gamma_2$ the spin dephasing, and $\Gamma_{GL}$ model the fast spin reset to the $\ket{0}$ state through the green light ($\lambda_L \sim 532$ nm) \cite{Doherty2013TheDiamond}. The Lindblad superoperator are defined by ${\cal L}[x_k]\rho=(2x_k\rho x_k^{\dagger}-x_k^{\dagger}x_k\rho-\rho x_k^{\dagger}x_k)$. 


\paragraph{Quantum Battery:-}
\begin{figure*}
    \centering
    \includegraphics[width=0.8\textwidth]{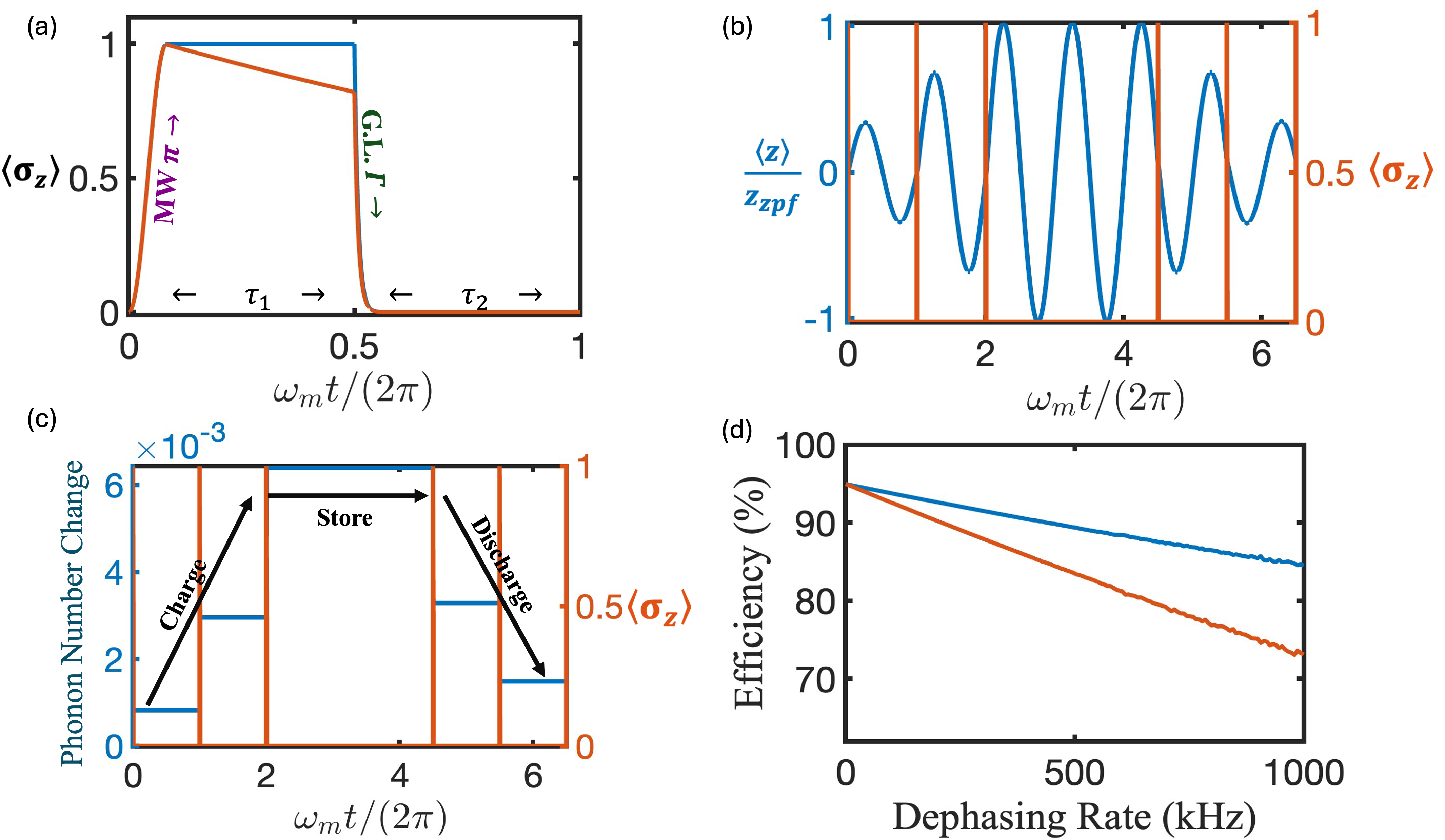}
    \caption{ {\it Quantum Battery Protocol:} Mechanical modes store energy coherently over long periods. The protocol cycles through initialization of spin, $\rho_i = \rho_m^i \otimes |0\rangle\langle 0|$, resonant a microwave $\pi$ pulse of Rabi frequency 5 MHz on the spin to couple spin and motion with coupling strength $g$=$2\pi\cdot3$ Hz, free evolution for $\Delta\tau = \tau_1$, spin reset to $\ket{0}$ via a green laser (decoupling spin and motion), and free evolution for $\Delta\tau = \tau_2$, where $\tau_1 + \tau_2 = T_m$, the mechanical period.
    \textbf{(a)} Expectation value $\langle \sigma_z \rangle$ during the control sequence; blue (without spin damping) and red (with spin damping at rate $\Gamma$) lines are shown.
    \textbf{(b)} With $\tau_1 = T_m/128$, the spin delivers brief "kicks" to the mechanics. Plotted are $\langle z \rangle / z_{\text{zpf}}$ (blue, left axis) and $\langle \sigma_z \rangle$ (red, right axis), starting from a thermal state with $\bar{n}_m = 2$. During {\it charging} (in-phase kicks at $\tau = 0, 1, 2$), the mechanical amplitude increases. In the {\it storage} phase (no kicks), the amplitude remains constant. For {\it discharging} (out-of-phase kicks at $\tau = 4.5, 5.5, 6.5$), the amplitude decreases. At $\tau = 6.5$, both spin and mechanics return to initial states.
    \textbf{(c)} Phonon number $\langle a^\dagger \rangle$ during the process, showing an increase during charging, constancy during storage, and decrease during discharging.
    \textbf{(d)} Charging (blue) and discharging (red) efficiencies as functions of spin dephasing rate. Spin damping (characterized by $T_1$) limits efficiencies, affecting discharging more than charging. }
    \label{fig:enter-label}
\end{figure*}
We now propose to use the spin-mechanical coupling to control the spin so as to store energy in the harmonic quantum motion of the nanodiamond. The motional modes can have very long lifetimes at low pressures and can effectively store the energy for extended periods, functioning as a battery. Quantum batteries that store energy in coherent quantum systems have recently garnered significant interest and spin systems coupled with harmonic oscillators are being explored as hybrid systems to store energy. 
Such systems typically use a harmonic oscillator mode to charge an ensemble of spins where the spin ensemble act as the battery \cite{Campaioli2024Colloquium:Batteries, Campaioli2017EnhancingBatteries, Zhang2019PowerfulBattery}. These studies required the transition energies of the spins and oscillator to be similar for resonant coupling, which can be practically challenging
\cite{Campaioli2024Colloquium:Batteries}.

In this section, we  study the NV-spin-mechanical-coupled system, and show that by driving the NV spin via 
pulses of microwave and green light, one can either {\it charge}, or {\it discharge}, the motional quantum mode - phonon mode - which acts as the battery, on demand. We do {\it not require} the transition energies of the spin and phonon modes to be similar, making the protocol applicable to cases where the phonon frequency can be more realistic $\omega_m/2\pi\sim {\rm Hz-MHz}$. The spin-mechanical battery protocol involves two key steps, as illustrated in FIG. 2(a):

\textbf{Step 1:} A fast resonant $\pi$ pulse transfers the spin from the ground state $\ket{0}$ to the excited state $\ket{-1}$. After this, the system is allowed to evolve for a time $\tau_1$. The pulse switches on the spin-mechanical interaction which generates a drive on the mechanics of the form $\mathcal{H}_{int}\sim  g(a+a^{\dagger})$, charging the phonon battery. 
In a realistic scenario, the spin will decay slightly towards a thermal state due to spin relaxation.

\textbf{Step 2:} At time $\tau_1$, a 532 nm green light pulse is then used to rapidly reset the spin to the $\ket{0}$ state, holding it there for a duration $\tau_2$, where  $\tau_1+\tau_2=2k\pi/\omega_m$. This reset operation is non-unitary and once reset, the mechanical drive is effectively switched off and the spin is in a product state with the mechanics.

We focus on the limit where $\tau_1\ll \tau_2$, as decoherence primarily occurs during $\tau_1$. To store energy in the battery, we begin with the mechanics in a thermal state and the spin in the $\ket{0}$ state e.g. $\rho(t=0)=\rho_{th}(\bar{n}_{th})\otimes\ket{0}\bra{0}$. We then repeatedly apply the battery protocol until the desired amount of energy is stored in the mechanical system. This energy storage is evidenced by a coherent displacement of the thermal state and a corresponding increase in the mean phonon number. To extract the stored energy, one can reverse the overall displacement of the thermal state by applying the control sequence on the spin in a time reversed manner relative to the original charging sequence. To be effective this reversed pulse sequence should be applied after a delay of $2(k+1)\pi/\omega_m$, following the completion of the charging pulses (FIG 2(b)-(c)).


For the lossless case, the charging and discharging pulse sequences should be symmetrical, and both the spin and phonon will return to a state identical to the initial state. The degree of motional charging is proportional to the duration $\tau_1$,  and the maximum charging over a single pulse occurs when $\tau_1=2\pi/\omega_m$, with the maximum phonon value achieved is $n_{max}=(g/2w_m)^2$ 
However, including decay of the spin, particularly dephasing, affects the charging and discharging efficiency, 
as seen in FIG. 2(d). Consequently, the system cannot completely return to its original mechanical state, leading to a loss of the stored energy. Therefore, the maximum amount of retrievable energy stored has an upper limit set by the decoherence processes in the spin system. Notably, the battery's operation is robust to the initial quantum state of the mechanics, eliminating the need for a specific initialization of the mechanical oscillator. 



\begin{figure}
    \centering
    \includegraphics[width=1.0\linewidth]{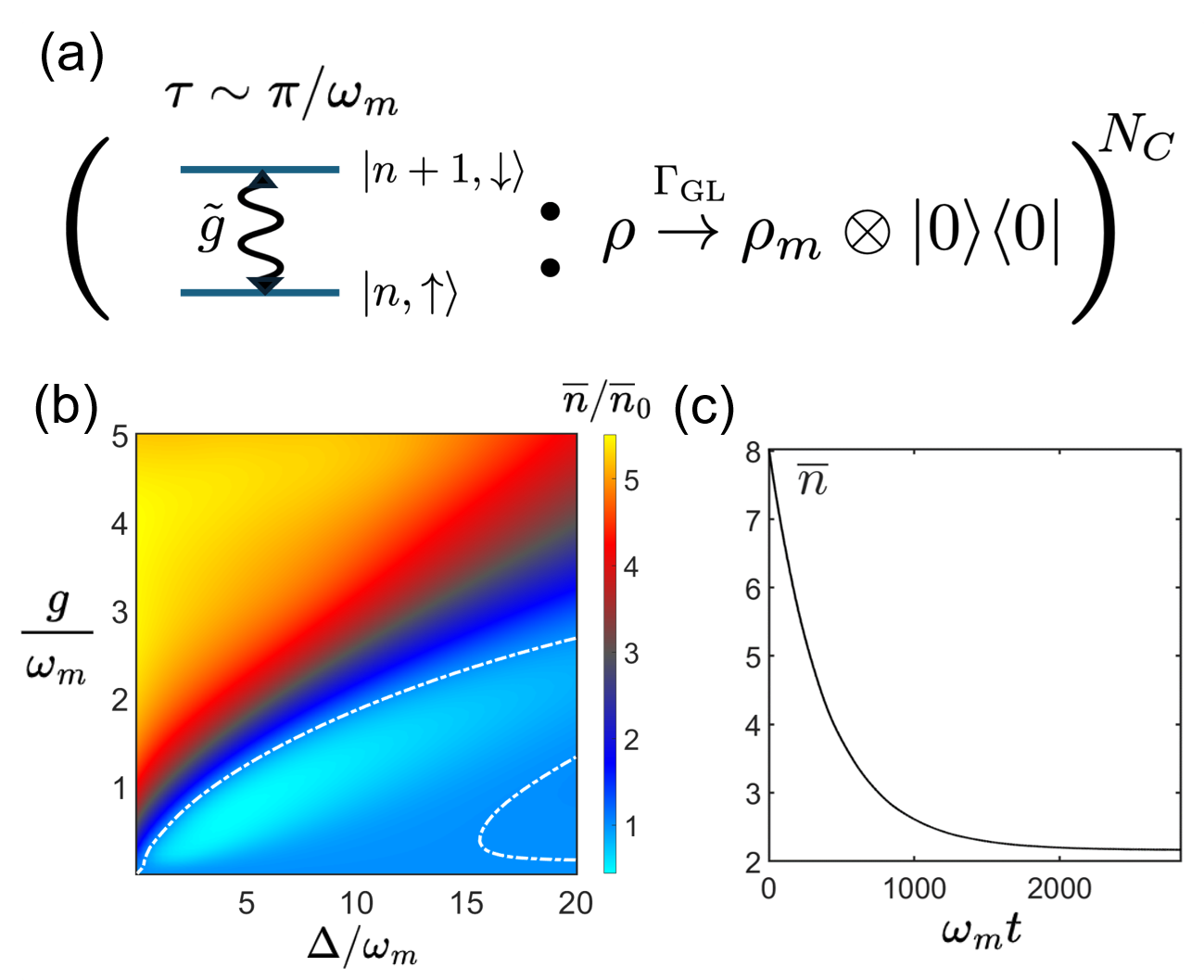}
    \caption{ \textbf{(a)} {\it Illustration of the Cooling Protocol:} The spin subsystem, coupled to a low-frequency mechanical oscillator with strength $\tilde{g}$, undergoes repeated operations. Each operation consists of a microwave drive (Rabi frequency $\Omega$) followed by a spin reset using green light at a rate $\Gamma_{\text{GL}}$. This sequence efficiently transfers energy from the mechanical mode to the spin, which is then cooled via the reset, reducing the mean phonon occupation number and cooling the mechanical system. The protocol is iterated $N_c$ times, with each iteration lasting $\tau \sim \pi/\omega_m$.
    \textbf{(b)} {\it Cooling Efficiency Map:} A 2D color map showing the ratio of the mean phonon number $\overline{n} = \langle a^\dagger a \rangle - \langle a^\dagger \rangle \langle a \rangle$ to its initial value $\overline{n}_0$ after $N_c = 100$ cooling cycles, plotted as a function of detuning $\Delta/\omega_m$ and coupling strength $g/\omega_m$. The system starts in a thermal state with $\overline{n}_0 = 8$. The white dashed line indicates where the ratio equals one. Light blue regions inside this contour highlight optimal cooling areas where $\overline{n} < \overline{n}_0$, demonstrating that precise parameter tuning enhances cooling.
    \textbf{(c)} {\it Time Evolution of Phonon Occupation:} The mean phonon number $\overline{n}(t)$ is plotted over time for the initial state $\rho(0) = \rho_{\text{th}}(\overline{n}_0) \otimes \ket{0}\bra{0}$ with $\overline{n}_0 = 8$. Parameters used are $(\Delta, g, \Omega, N_c) = (2\omega_m, \omega_m/2, \omega_m/2, 900)$; other parameters are listed in Table I. }
    \label{fig:cool}
\end{figure}

\paragraph{Cooling:-}\label{coolsection}
In the previous section we described a pulse sequence to displace the phonon state in phase space in a controlled manner but the purity of the motional state was not altered.  Here we show that one can change the purity of the phonon state using a particular sequence of pulses on the spin. In particular one can cool the mechanical oscillator, reducing its mean phonon occupation number $\langle a^{\dagger}a\rangle$. This cooling can be interpreted as interaction with an effective cold bath at temperature $T_c$. To show this we start with the spin-mechanical system described by \eqref{Hspin2}, and move to a displaced frame given by $U\left(\theta\right)D\left(g/\omega_m\right)$, where $D\left(g/\omega_m\right)=e^{(g/\omega_m)\left(a^{\dagger}-a\right)}$, and  $U\left(\theta\right)=\exp\left(-i(\theta/2)\sigma_y\right)$ with $\theta=\arctan\left(-2\Omega/\delta\right) $ and $\delta=\Delta-2g^2/\omega_m$. The Hamiltonian \eqref{Hspin2} transforms to  
\begin{equation}\label{dressH}
    \overline{\mathcal{H}}=\omega_m a^{\dagger}a-\frac{\overline{\Delta}}{2}\sigma_z+\frac{g}{2\sqrt{1+\left(2\Omega/\delta\right)^2}}\left(\sigma_z+\frac{2\Omega}{\delta}\sigma_x\right)\left(a+a^{\dagger}\right)
\end{equation}
where $\overline{\Delta}=\sqrt{\delta^2+4\Omega^2}$ and where the new dressed basis is expressed as 
\begin{align}
    \ket{\downarrow}&=-\sin{\frac{\theta}{2}}\ket{0}+\cos{\frac{\theta}{2}}\ket{-} \\
    \ket{\uparrow}&=\cos{\frac{\theta}{2}}\ket{0}+\sin{\frac{\theta}{2}}\ket{-}.
\end{align} In this frame, the interaction has two components: a longitudinal coupling, ${\cal H}_\parallel\sim\sigma_z\left(a + a^{\dagger}\right)$, which results in spin-mechanical Zeeman splitting, and a transverse coupling, ${\cal H}_\perp\sim\sigma_x\left(a + a^{\dagger}\right)$, which leads to energy exchange between the spin and the mechanical modes. 
The Jaynes-Cummings Hamiltonian is obtained from \eqref{dressH} assuming $\vert\omega_m-\overline{\Delta}\vert\ll \overline{\Delta}+\omega_m,\,\omega_m$ and after neglecting the fast rotating terms using the rotating wave approximation (RWA).
\begin{equation}\label{HJC}
    \mathcal{H}^{\text{eff}}=\omega_m a^{\dagger}a-\frac{\overline{\Delta}}{2}\sigma_z+\Tilde{g}\left(a\sigma_++a^{\dagger}\sigma_-\right)
\end{equation}
with $\Tilde{g}=\left(g\Omega/\delta\right)/\left(\sqrt{1+\left(2\Omega/\delta\right)^2}\right)$. As a result, when the spin is pulsed in the original frame \eqref{Hspin2}, the Hamiltonian in the displaced dressed frame \eqref{dressH} is equivalent to a beam-splitter interaction where energy can be transferred between the phonon and spin. 
To efficiently cool the mechanical oscillator, we periodically apply a pulse of green light to reset the spin after a period of beam-splitter interaction.  This fast reset process is described by the master equation \eqref{be} with the Lindblad superoperator ${\cal L}_{\Gamma_{GL}}[\sigma_-]$. We illustrate the cooling process with a diagram in FIG. \ref{fig:cool}(a).

Given that the spin's energy levels are in the MHz-GHz range—significantly higher than the mechanical frequency—we must bring the spin and mechanical oscillator into resonance for efficient energy transfer. This is achieved by driving the spin in the initial frame \eqref{Hspin2} with a Rabi frequency $\Omega$, dressing the spin's energy levels to arrange that the dressed level transition frequency aligns with the mechanical frequency $\overline{\Delta} \sim \omega_m$. The result is an effective beam splitter Hamiltonian as described by \eqref{HJC}. This interaction facilitates the efficient transfer of energy between the mechanical oscillator and the spin. After an interaction time $t_f\sim \pi/\omega_m$, we reset the spin to its ground state $\ket{0}$ using a green laser pulse. This process dissipates the energy gained by the spin into the environment, thereby cooling the mechanical oscillator incrementally with each cycle. This protocol is repeated $N_C$-times till the phonon occupation number $\overline{n}=\langle a^\dagger a\rangle-\langle a^\dagger\rangle\langle a\rangle$ reaches a stationary value. 

Ideally, resetting the spin to the ground state of the dressed basis will lead to optimal cooling. However, the spin's coherence time $T_{2}^*$ poses a limitation, as the states ${\ket{\downarrow},\ket{\uparrow}}$ are tilted in the Bloch sphere, making them susceptible to dephasing. For slow mechanical oscillators, this limitation could hinder cooling efficiency. To mitigate this, we do not initialize the spin to the ground state of the dressed basis but we initialize the spin in the $\ket{0}$ state and then apply the dressing drive. 

To determine the optimal parameters for cooling, we numerically sweep the detuning $\Delta$ and coupling strength $g$. FIG. \ref{fig:cool}.(b) reveals a threshold for the coupling strength, beyond which cooling becomes ineffective, irrespective of the detuning. Effective cooling is achieved at low coupling and detuning values, as evidenced by the significant reduction in the phonon number.

{\it Quantum Otto cycle:} The traditional Otto cycle is a fundamental thermodynamic cycle used to describe the operation of classical heat engines. It consists of a four-stroke cycle made of adiabatic expansion, isochoric cooling, adiabatic compression, and isochoric heating. We propose to adapt the Otto cycle to a quantum mechanical framework by considering the low frequency mechanical oscillator presented in FIG. \ref{fig:Qbat} as the working fluid - the phonons. In traditional heat engines, the heating and cooling parts of the cycle are typically achieved by connecting, in a well-controlled way, the working fluid with two thermal baths at two temperatures $T_h >T_c$. Instead, we consider the working fluid, the phonons of the mechanical oscillator, to be in constant thermal contact with the hot bath (at the rate $\gamma_m$), and will apply the spin-mechanical cooling protocol when we wish this working fluid to be in contact with the cold bath. 
Referring to FIG. 4(a), the four-stroke of the cycle can then be defined as the following

\textbf{Adiabatic expansion $1\rightarrow 2$} The system starts out in a product state of $\rho_1=\rho_{th}(\bar{n}_{th})\otimes\ket{0}\bra{0}$, where $\bar{n}_{th}$ is the mean phonon occupation when in thermal equilibrium with the hot bath. During the adiabatic expansion phase, the mechanical oscillator's frequency is decreased smoothly from its initial value $\omega_m$ to a lower value over a time period $T$. We consider a linear expansion $\omega(t)=\omega_m\left(1-\lambda \frac{t}{T}\right)$ where $\lambda=\left(d_t\omega\right) \frac{T}{\omega_m}<1$ and $\left(d_t\omega\right)$ represents the rate of frequency change. The work done during this stage  is $\langle W_e \rangle =\langle H_2\rangle-\langle H_1\rangle<0$, where $\langle H_1\rangle$ and $\langle H_2\rangle$ are the mean energies of the system at the beginning and end of the expansion, respectively. 

\textbf{Cold isochore $2\rightarrow 3$} In this phase, the system is cooled at the fixed mechanical frequency $\omega_m\left(1-\lambda\right)$ by the protocol presented in the previous section to reach an effective temperature $T_c\equiv T_3$, which can be preset by choosing values $(g,\Delta)$, see FIG. \ref{fig:cool}(b). 
The heat released by the working fluid during this stage is  
$\langle Q_c\rangle =\langle H_3\rangle-\langle H_2\rangle$, where $\langle H_3\rangle$ is the mean energy after cooling. 

\textbf{Adiabatic compression $3\rightarrow 4$} Following cooling, the mechanical oscillator undergoes compression. The frequency is increased from $\omega_m\left(1-\lambda\right)$ to $\omega_m$ over the same time period $T$, following the trajectory: $\omega(t)=\omega_m\left[1-\lambda\left(1-\frac{t}{T}\right)\right]$. The work done on the working fluid is $\langle W_c \rangle =\langle H_4\rangle-\langle H_2\rangle>0$, where $\langle H_4\rangle$ is the energy at the end of the compression. 

\textbf{Hot isochore $4\rightarrow 1$} The mechanical oscillator is uncoupled from the spin and freely interacts with the thermal environment, absorbing heat until it reaches, eventually, the initial thermal state where $\langle a^\dagger a \rangle =n_{th}$. The heat exchange is given by $\langle Q_h\rangle =\langle H_1\rangle-\langle H_4\rangle$. This stage completes the cycle, restoring the system to its initial state and making it ready for the next cycle. 

\begin{figure}
    \centering
    \includegraphics[width=1.05\linewidth]{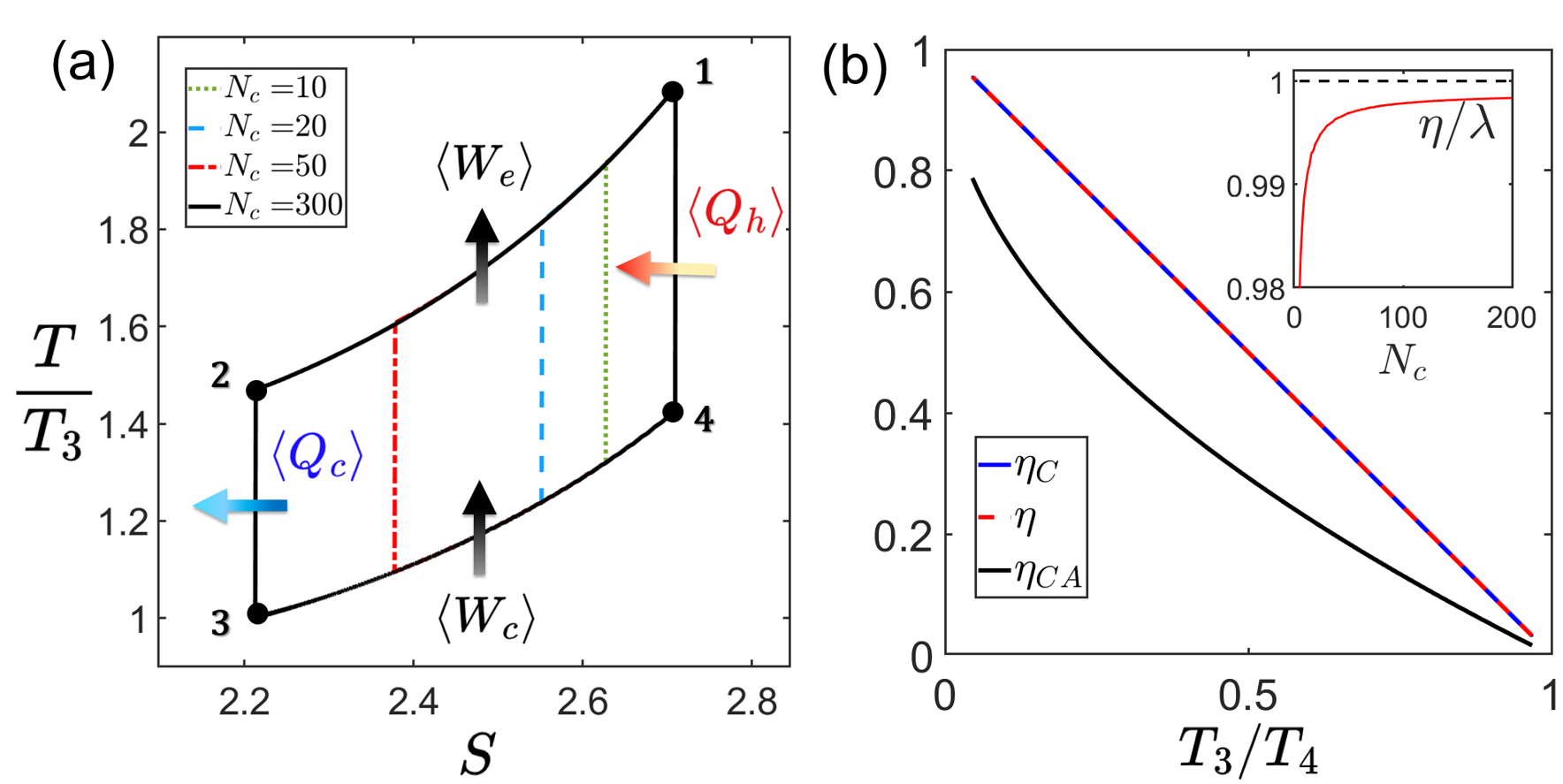}
    \caption{ \textbf{(a)} Temperature-entropy (T-S) diagram of the quantum Otto cycle for various numbers of cooling cycles $N_c$. The temperature and entropy at each point are calculated using $\beta_k \hbar \omega_k = \ln\left(\frac{\overline{n} + 1}{\overline{n}}\right)$ and the von Neumann entropy $S(\rho) = -\text{Tr}[\rho \ln \rho]$, respectively. The diagram illustrates the heat exchanged with the working fluid ($Q_c$, $Q_h$) and the work done by and on the system ($W_e$, $W_c$). Parameters used are $g = 0.25,\omega_m$, $T = \pi/\omega_m$, $\Delta = 3,\omega_m(1 - \lambda)$, and $d_t\omega = 10^4,\text{rad/s}^2$.
    \textbf{(b)} Comparison of the theoretical Carnot efficiency (solid blue line), the observed efficiency of our quantum Otto heat engine (dashed red line, see Eq.~\eqref{realeff}), and the Curzon–Ahlborn efficiency \cite{curzon1975} (solid green line) as functions of the temperature ratio $T_3/T_4$. All parameters are fixed except the compression/expansion time $\omega_m T$, which is varied from $3\pi$ to $10\pi$, resulting in different values of $\lambda$ (with $\lambda \equiv \eta_C$) and $T_3/T_4$. The initial phonon occupation number is set to 7, and the cooling protocol is applied 250 times, with each cooling step lasting $\pi/\omega_m$. We use $d_t\omega = 10^4\,\text{rad/s}^2$. The inset shows the ratio of the engine's computed efficiency (Eq.~(13)) to the Carnot efficiency as a function of the number of cooling cycles up to $N_c = 200$, with $\omega_m T = \pi/10$. The engine achieves near-optimal efficiency even with a low number of cooling cycles. Other parameter values are listed in Table I.   }
    \label{fig:QCycle}
\end{figure}

In FIG. \ref{fig:QCycle}(a), we plot the temperature-entropy (T-S) diagram of the engine for different values of cooling cycle $N_c$. The temperature and entropy at each point along the cycle are computed from $\beta_k \hbar\omega_k=\log\left(\frac{\overline{n}_k+1}{\overline{n}_k}\right)$, and $S(\rho_k)=-\rho_k\,\log\rho_k$, where $\overline{n}_k$ and $\rho_k$ are respectively the phonon occupation number and density matrix at the $k$-point. We note that $N_c$ is increased the cycle can reach lower $S$ values as the effective cold bath temperature $T_c$ is decreased. The form of the cycle is similar to the classical Otto cycle from which four temperatures are defined. To quantize the efficiency of the heat engine, we defined, similarly to classical heat engines, the quantum efficiency $\eta$ of the cycle. It is defined as the ratio of net work output to the heat absorbed from the hot bath: 
\begin{equation}\label{realeff}
    \eta=1+\frac{\langle Q_c\rangle}{\langle Q_h\rangle}=-\frac{\left(\langle W_e\rangle+\langle W_c\rangle\right)}{\langle Q_h\rangle}\;\;.
\end{equation}
In analogy to the maximum Carnot efficiency, for an ideal Otto cycle, the theoretical quantum efficiency limit can be expressed in terms of the frequencies
\begin{equation}
    \eta_{C}=1-\frac{\omega_2}{\omega_1}=\lambda\;\;,
\end{equation}
where $\omega_1=\omega_m$ and $\omega_2=\omega_m\left(1-\lambda\right)$. In FIG. \ref{fig:QCycle}.(b), we compare, as a function of $T_3/T_4$, the efficiency $\eta$ obtained for the engine with the Carnot limit $\eta_c\equiv\lambda$ and the Curzon–Ahlborn efficiency $\eta_{CA}$, 
which quantifies the efficiency of a heat engine operating at maximum power \cite{curzon1975}. 
This spin-mechanical quantum heat engine demonstrates an efficiency approaching Carnot efficiency without the need for quasi-static operation. The low phonon bath coupling rate $\gamma_m$,  plays a crucial role in this, as it minimize energy dissipation and associated irreversibilities allows the engine to exceed the Curzon–Ahlborn efficiency. 
In the inset of FIG. \ref{fig:QCycle}(b), we  plot the ratio $\eta/\lambda$ for different numbers of cooling cycles $N_c$. It shows that even for $N_c\sim 1$, the engine operates very close to the maximum Carnot efficiency.

\paragraph{Acknowledgements--}The authors would like to thank Kani Mohamed and Thomas Fogarty for their insightful discussions. This work was supported by funding from the Okinawa Institute of Science and Technology Graduate University. We are grateful for the help and support provided by the Scientific Computing and Data Analysis section of Core Facilities at OIST.
\\
$^\dagger$ M.H. and A.N. contributed equally to this work.

\begin{table}[h]
\begin{center}
\caption{Parameters used in this paper
\label{table:params}}
\begin{threeparttable}
\begin{tabular}{p{4cm}p{2cm}p{2cm}}
\hline\hline
  Quantity & Value & Unit   \\ \hline
 Mechanical frequency $\omega_m$  & ($2\pi\cdot 50$) & rad/s \\
 Quality factor Q & (10000) & - \\
 Dephasing time $T_2^*$ & $10^{-3}$ & s \\
 Thermal relaxation $T_1$ & $2\cdot10^{-3}$ & s \\
 Initialisation rate $\Gamma_L$ & $10^{5}$ & Hz \\
 Initialisation time $t_{lase}$ & $5\cdot10^{-5}$ & s\\
\hline
\end{tabular}

\end{threeparttable}
\end{center}
\vspace{-7mm}
\end{table}

\input{main.bbl}

\end{document}

%% file: main.bbl
%